\documentstyle[twocolumn,aps]{revtex}
\input epsf

\newcommand{\pp}{\partial }
\newcommand{\ww}{\mbox{\tiny $\wedge$}}
\newcommand{\spc}{\mbox{\hspace{1cm}}}

\begin{document}

\title{Anti-de Sitter space and black holes}

\author{M\'aximo Ba\~nados$^{1,2,3}$, Andr\'es Gomberoff$^{2}$ and
Cristi\'an Mart\'{\i}nez$^{2}$ }

\address{$^1$Departamento de F\'{\i}sica Te\'orica, Universidad de
Zaragoza, Ciudad Universitaria 50009, Zaragoza, Spain, \\
$^2$Centro de Estudios Cient\'{\i}ficos de Santiago, Casilla 16443,
Santiago 9, Chile, \\ 
$^3$Departamento de F\'{\i}sica, Universidad de
Santiago de Chile, Casilla 307, Santiago 2, Chile}

\maketitle

\begin{abstract}
Anti-de Sitter space with identified points give rise to black-hole
structures. This was first pointed out in three dimensions, and generalized to higher dimensions by Aminneborg et al. In this paper, we analyse
several aspects of the five dimensional anti-de Sitter black hole
including, its relation to thermal anti-de Sitter space, its embedding
in a Chern-Simons supergravity theory, its global charges and holonomies, and the existence of Killing spinors.   
\end{abstract}

\section{Introduction}

Despite of the ``unphysical" properties of coupling a negative
cosmological constant to general relativity, anti-de Sitter space has
a number of good properties. It has been shown to be stable
\cite{Abbot-Deser}, and to possess positive energy representations
\cite{Fronsdal} (see \cite{Gibbons} for a review of further properties
of anti-de Sitter space). Recently, anti-de Sitter space has
appeared in a surprising new context. Maldacena \cite{Maldacena}
has conjectured  that the large $N$ limit of certain super conformal
theories are equivalent to a string theory on a background containing
the direct product of anti-de Sitter space (in five dimensions) times
a compact manifold. The conjecture raised in \cite{Maldacena} has been
interpreted as an anti-de Sitter holography in \cite{Witten98} by
noticing that the boundary of anti-de Sitter space is conformal
Minkowski space. In this sense, the relation between conformal field
theory in four dimensions and type IIB string theory on adS$_5\times
M_5$ is analog to the relation between three dimensional Chern-Simons
theory and 1+1 conformal field theory \cite{Witten89}.   

The relationship between 5D anti-de Sitter space and conformal field theory is particularly interesting in the context of quantum black holes. The reason is that, in 2+1 dimensions, the adS/CFT correspondence has provided interesting proposals to understand the 2+1 black hole entropy \cite{Carlip,Strominger}. One may wonder whether there exists black holes analogous to the 2+1 black hole in five dimensions. This is indeed the case. It is now known that identifying point in anti-de Sitter space, in any number of dimensions, gives rise to black hole structures.  These black holes, often called ``topological black holes", were first discussed in three dimensions in \cite{BTZ,BHTZ}, in four dimensions in \cite{O}, and in higher dimensions in \cite{BTZu,b3}. Further properties have been studied in \cite{tbh}.

A topological black hole can be defined as a spacetime whose local
properties are trivial (constant curvature, for example) but its
causal global structure is that of a black hole.  The higher
dimensional ($D>3$) topological black hole has an interesting causal
structure displayed by a $D-1$ Kruskal (or Penrose-Carter) diagram, as
opposed to the usual 2-dimensional picture \cite{b3}. Indeed, the
metric in Kruskal coordinates has an explicit $SO(D-2,1)\times SO(2)$
invariance, as opposed to the Schwarzschild black hole having an
$SO(1,1)\times SO(D-1)$ invariance. Accordingly, the topology of this 
black hole is $M^{D-1}\times S_1$ with $S_1$ being the
compactified coordinate, and $M^n$ denotes $n-$dimensional conformal
Minkowski space.  Just as for ordinary black holes, in the Euclidean
formalism the Minkowski factor becomes $\Re^{D-1}$.  This topology has
to be compared with $M^2\times S_{D-2}$ arising in usual situations
like the $D$-dimensional Schwarzschild black hole. 

We shall start (Sec. II) by reviewing the main properties of the four
dimensional topological black hole. Of particular interest is the
Euclidean black hole obtained by identifications on Euclidean anti-de
Sitter space. We shall exhibit in section \ref{Euc/4} the explicit
relation between the Euclidean manifold considered in \cite{Witten98},
as the Euclidean sector of the topological black hole. We then
introduce the five dimensional black hole metric (Sec. III), and study
its embedding in a Chern-Simons theory in Sec. IV. The black hole has
a non-trivial topological structure and its associated  holonomies are
calculated in Sec. V.    Finally, we shall consider other possible
solutions with exotic topologies following from an ansatz suggested by
the topological black hole in $D$ dimensions. 

\section{The four dimensional topological black hole}

We shall start by considering the four dimensional case. For a
exhaustive analysis and classification of the properties of anti-de
Sitter space with identifications see \cite{Holst-Peldan}. Here we
shall review and further develop the three dimensional causal
structure introduced in \cite{b3} and, in particular, the form of the
metric in Kruskal coordinates.   

\subsection{Identifications and causal structure}

In four dimensions anti-de Sitter space is defined as the universal
covering
of the surface
\begin{equation}
-x_0^2 + x^2_1 + x_2^2 + x^2_3 - x_4^2  = -l^2.
\label{ads4}
\end{equation}
This surface has ten Killing vectors or isometries and define the
anti-de Sitter group. 

We shall identify points on this space along the boost 
\begin{equation}
\xi =\frac{r_+}{l} (x_{3} \partial_{4} + x_4 \partial_{3}), \ \ \
\xi^2=\frac{r_+^2}{l^2}(-x_{3}^2+x_4^2),
\end{equation}
where $r_+$ is an arbitrary real constant. The norm of $\xi$ can be
positive, negative or zero. This is the key property leading to black
holes \cite{BTZ,BHTZ}. Identifications along a rotational Killing
vector produce the conical singularities discussed in \cite{DJt} which
do not exhibit horizons.

In order to see graphically the role of the identifications, we  plot
parametrically the surface (\ref{ads4}) in terms of the values of
$\xi^2$.  
\begin{figure}
\begin{center} 
\epsfxsize=7cm
  \leavevmode
\epsfbox{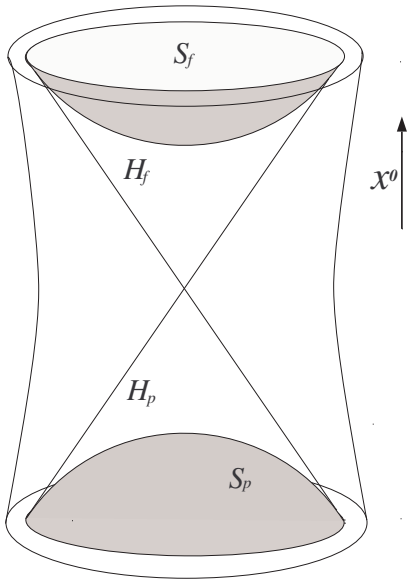}\label{f1}
\vskip -3cm
\caption{Anti-de Sitter space. Each point in the figure
represent the pair ${x_3,x_4}$ with $x_3^2-x_4^2$ (the norm of the Killing vector) fixed. }
\end{center}
\end{figure}
First we consider the surface $\xi^2=r_+^2$ giving rise to the null
surface
\begin{equation}
x_0^2=x_1^2 +  x_2^2.
\label{cone}
\end{equation}
Then we consider the surface $\xi^2=0$ which is represented by the
hyperboloid
\begin{equation}
x_0^2=x_1^2 + x_{2}^2 + l^2.
\label{hyper}
\end{equation}
In Fig. 1, the `time' coordinate $x_0$ is drawn along the z-axis.
Hence, future directed light cones are oriented upwards.  
The cone (\ref{cone}) has two pointwise connected branches, called
$H_f$ and $H_p$, defined by
\begin{eqnarray}
H_f : \spc x^0=+\sqrt{x_1^2 +  x_2^2} \,,\\
H_p : \spc x^0=-\sqrt{x_1^2 +  x_2^2}\,.
\end{eqnarray}
Similarly, the hyperboloid (\ref{hyper}) has two disconnected branches
that we have called $S_f$ and $S_p$,
\begin{eqnarray}
S_f : \spc x^0=+\sqrt{l^2 + x_1^2 + x_2^2}\,,\\
S_p : \spc x^0=-\sqrt{l^2 + x_1^2 + x_2^2}\,.
\end{eqnarray}

The Killing vector $\xi$ is spacelike in the region contained in-between $S_f$ and $S_p$, it is null at $S_f$ and $S_p$, and it is timelike in the causal future of $S_f$ and in the causal past of $S_p$.

We now identify points along the orbit of $\xi$.  The 1-dimensional manifold orthogonal to Fig. 1 becomes compact and isomorphic to $S_1$.  The region where $\xi^2<0$ becomes chronologically pathological because the identifications produce timelike curves. It is thus natural to remove it from the physical spacetime.  The hyperboloid is thus a singularity because timelike geodesics end there. In that sense $S_f$ and $S_i$ represent the future and past singularities, respectively. 

Once the singularities are identified one can now see that the upper
cone $H_f$ defined in (\ref{cone}) represents the future horizon.
Indeed, $H_f$ coincides with the boundary of the causal past of
lightlike infinity. In other words, all particles in the causal future
of $H_f$ can only hit the singularity.  Because infinity is connected
in this geometry, as opposed to the usual Schwarzschild black hole
having two disconnected asymptotic regions, the lower cone does not
represent a horizon in the usual sense. Note, however, that the lower
cone does contain relevant invariant information. For example, all
particles in the causal past of $H_p$ come from the past singularity.   

In summary, we have seen that by identifying points in anti-de Sitter
space one can produce a black hole with an unusual topology. The
causal structure is displayed by a three dimensional Penrose diagram
(see fig 2), as opposed to the usual two dimensional picture. Since,
each point in Fig. 1  represents a circle, the topology of this black
hole is $M^3\times S_1$, where we denote by $M^n$ conformal Minkowski
space in $n$ dimensions.    
\begin{figure}
\begin{center} 
\epsfxsize=7cm
  \leavevmode
 \epsfbox{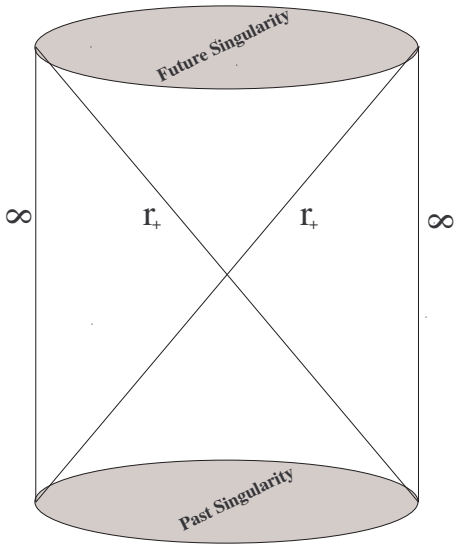}
\label{f2}
\caption{Penrose Diagram}
\end{center}
\end{figure}

\subsection{Kruskal coordinates}

So far we have not displayed any metric. The $M^3\times S_1$ black
hole can be best described in terms of Kruskal coordinates.  Let us
introduce local coordinates on anti-de Sitter space (in the region
$\xi^2>0$) adapted to the Killing vector used to make the
identifications. We introduce the $4$ dimensionless local coordinates
$(y_\alpha,\varphi)$ by
\begin{eqnarray}
x_\alpha&=&\frac{2l y_\alpha}{1-y^2},  \spc  \alpha=0,1,2 \nonumber \\
x_3&=&\frac{lr}{r_+}  \sinh\left(\frac{r_+\varphi}{l}\right),
\label{K}\\ 
x_4&=&\frac{lr}{r_+}  \cosh\left(\frac{r_+\varphi}{l}\right),
    \nonumber 
\end{eqnarray}
with 
\begin{equation}
r = r_+ \frac{1+y^2}{1-y^2}, \ \ \ \ y^2 = -y_0^2 +y_1^2 +y_2^2.
\label{r-y}
\end{equation}
Before the identifications are made, the coordinate ranges are
$-\infty < \varphi < \infty$ and $-\infty < y^\alpha <\infty$ with the
restriction $-1<y^2<1$. Note that the boundary $r\rightarrow\infty$
correspond to the hyperbolic ``ball" $y^2=1$. The induced metric has
the Kruskal form
\begin{equation} 
ds^2 =  \frac{l^2(r+r_+)^2}{r_+^2}\, (-dy_0^2 + dy_1^2 +dy_2^2)  + r^2
d\varphi^2, 
\label{ds/krus}
\end{equation}
and the Killing vector reads $\xi =\partial_\varphi$ with
$\xi^2=r^2$. The
quotient space is thus simply obtained by identifying $\varphi \sim
\varphi+2\pi n$, and the resulting topology clearly is $M^3\times
S_1$. With the help of (\ref{K}), it is clear that the Kruskal diagram
associated to this geometry is the one shown in Fig. 1. Thus, the
metric (\ref{ds/krus}) represents the $M^3\times S_1$ black hole
written in Kruskal coordinates.   

The metric (\ref{ds/krus}) has some differences with the usual
Schwarzschild--Kruskal metric that are worth mentioning. First of all,
infinity is connected. There is only one asymptotic region with the
topology of a cylinder $\times S_1$ and thus, in particular, only one
patch of Kruskal coordinates is needed to cover the full black hole
spacetime. Second, the metric has an explicit $SO(2,1)\times SO(2)$
symmetry. The presence of the $SO(2,1)$ factor is a consequence of the
three dimensional character of the causal structure. One could project
down the Penrose diagram (Fig. 2) to a flat diagram, as done in
\cite{O,Holst-Peldan}, but this is not natural and makes the causal
structure more complicated.

The issue of trapped surfaces in the four dimensional topological
black hole was studied in \cite{Holst-Peldan} where it was proved that
there exists a non-trivial apparent horizon.  We would like to point
out here that if one defines trapped surfaces as suggested by the
three dimensional Kruskal diagram, then the apparent horizon is not
present. 

We shall say that a closed spatial surface is trapped if its evolution
along any light like curve diminishes its area.  Conversely, we say
that the surface is not trapped if there exists at least one light
like curve along which the surface increases its area. Since every
point in Fig. 1 represents a circle, trapped surfaces are circles in
this black hole and not spheres or tori. 

The existence of trapped surfaces (as defined above) in the black hole
geometry can be easily checked using the Kruskal coordinates. At each
point in Fig. 1 labeled by coordinates $(y^0,y^1,y^2)$ there is a
circle of radius $r(y^{\alpha})$, and inside the region $H_{f}$ of
Fig. 1, $y^2 < 0$ and $y^0 > 0$. From (\ref{r-y}) we see that a
variation in the coordinates $y^{\alpha} (\alpha=0,1,2)$ produces the
following variation in $r$ 
\begin{equation}
\delta r(y^{\alpha}) = \frac{2r_+}{(1-y^2)^2}y_{\alpha}\delta
y^{\alpha}\ .
\label{tr}
\end{equation}
Since we move along a light like curve $(\delta y)^2 =0$. It is then
easy to see that the product $y_{\alpha}\delta y^{\alpha}$ is negative
(just take a basis where $y^{\alpha}=(1,0,0)$). Thus in the region
which is at the causal future of the horizon (inside $H_{f}$) the
circles are trapped. 

In the same way, one can check that the region outside the horizon has
non-trapped surfaces, and the region in the causal past of the past
horizon is inversely trapped: all surfaces increase their areas. 

\subsection{Schwarzschild metric}

We have seen that anti-de Sitter spacetime leads naturally to the
existence of a $M^3\times S_1$ black hole, and we have found an
explicit form for the metric in Kruskal form.  A natural question now
is whether one can find a global set of coordinates for which the
metric takes Schwarzschild form. We shall now prove that locally one
can find such coordinates but they do not cover the full manifold, not
even the full outer region.  Indeed, the black hole spacetime is not
static.  

This has a close analogy with the Schwarzschild black hole situation.
In that case, the spherically symmetric coordinates cover the outer
manifold and the metric looks static there. The maximal extension
covers the full manifold but it is not static. In our case, the
maximal extension is not static and covers the full manifold (and only
one patch is necessary because infinity is connected). The spherically
symmetric coordinates cover only part of the spacetime (not even the
full outer region) but the metric looks static.  

We introduce local ``spherical" coordinates $t,\theta$ and $r$ in the
hyperplane $\{y_0,y_1,y_2\}$:
\begin{eqnarray}
y_0 &=& f \sin\theta \sinh(r_+t/l), \nonumber\\
y_1 &=& f \sin\theta \cosh(r_+t/l), \nonumber\\
y_2 &=& f \cos\theta,  \label{schw1} 
\end{eqnarray}
with $f(r) = [(r-r_+)/(r+r_+)]^{1/2}$ and the ranges $0 < \theta<\pi$,
$-\infty<t<\infty$ and $r_+<r<\infty$.  The metric (\ref{ds/krus})
acquires
the Schwarzschild form  
\begin{equation}
ds^2 = l^2 N^2 d\Omega_2 +  N^{-2} dr^2 + r^2 d\varphi^2, 
\label{ds/sch}
\end{equation}
with $N^2(r) = (r^2 -r_+^2)/l^2 $, and
\begin{equation}
d\Omega_2 = - \sin^2{\theta}\, dt^2 + \frac{l^2}{r_+^2} d\theta^2 
\label{sph2}
\end{equation}
is the arc length of a hyperbolic 2-sphere. In these coordinates the
horizon is located at $r=r_+$, the point where $N^2$ vanishes.  

In the coordinates $(t,\theta,r,\varphi)$, the metric looks static and
has a Schwarzschild form, however, they do not cover the full outer
region. Indeed, the difference $y_1^2 - y_0^2$ is constrained to be
positive in the region covered by these coordinates.  This has to be
expected, because  it has been proved in \cite{Holst-Peldan} that
there are no timelike, globally defined, Killing vector in this
geometry. Note that $\partial/\partial t$ is a timelike Killing vector
of (\ref{ds/sch}) with norm $-(y_1^2 - y_0^2)$. 

It is also clear that one cannot find Schwarzschild coordinates
interpolating the outer and inner regions. For $r<r_+$, the metric
(\ref{ds/sch}) changes its signature and therefore it does not
represent the interior of the black hole.   Finally note that the
metric in the form (\ref{ds/sch}) has some similarities with the metric used in \cite{Maldacena} in the adS/CFT correspondence,
\begin{equation}
ds^2 = \frac{dU^2}{U^2} + U^2 (-dx_0^2+dx_1^2+dx_2^2+dx_3^2) \,.
\label{U}
\end{equation}
The function $U^2$ is multiplying four coordinates rather than a
single one, thus the point $U=0$, called the horizon, has the higher
dimensional structure described above. To pass to the region $U<0$ one
would probably  need to make an analytic extension (Kruskal extension)
of (\ref{U}).

There is another change of coordinates of (\ref{ds/krus}), which has
the advantage of covering the entire exterior of the Minkowskian black
hole geometry:  
\begin{eqnarray}
y_0 &=& f \sinh(r_+t/l), \nonumber\\ 
y_1 &=& f \cos\theta \cosh(r_+t/l), \nonumber\\
y_2 &=& f \sin\theta\cosh(r_+t/l) , 
\end{eqnarray} 
where $0 < \theta<2\pi$, $-\infty<t<\infty$ and $r_+<r<\infty$.
As is clear from the discussion above, the metric in this coordinate
frame must show a explicit dependence in time. In fact,
\begin{equation}
ds^2=N^2 l^2 d\Omega + N^{-2}dr^2 + r^2 d\varphi^2 ,
\label{ds2/sch}
\end{equation}
with $d\Omega = -dt^2 + (l^2/r_+^2)\cosh\left(r_+t/l \right)d\theta^2$
Note that for any constant value of $\theta$, this metric describes a
three dimensional black hole \cite{BTZ}.

\subsection{The vacuum solution}

We shall now study the metric in the limit $r_+\rightarrow 0$. Since
in this limit the Killing vector is not well defined, one may wonder
whether the vacuum state $r_+=0$  corresponds to anti-de Sitter space,
or anti-de Sitter space with identified points. In 2+1 dimensions the
vacuum has identified points and the corresponding Killing
vector belongs to a different class in the classification of all
possible isometries \cite{BHTZ}. 

The metric (\ref{ds/sch}) does not have a smooth limit for $r_+\rightarrow 0$. In order to take the limit we need to redefine the coordinate $\theta$,
\begin{equation}
\tilde \theta = \frac{l}{r_+}\theta \ . 
\end{equation}
Now we can take the limit $r_+\rightarrow 0$ in (\ref{ds/sch}) and,
after defining $\rho=l/r$, the metric takes the final form
\begin{equation}
ds^2_{r_+=0}=\frac{l^2}{\rho^2} ( -dt^2 + d\tilde\theta^2 + d\rho^2 +
d\varphi^2)
\label{v}
\end{equation}
where, in the limit $r_+=0$, the range of $\tilde\theta$ becomes
$0<\tilde\theta<\infty$. 

In this form it is clear that this space represents anti-de Sitter
with identified points. Locally, (\ref{v}) is isomorphic the
upper-half Poincar\'e plane and therefore it does have constant
curvature. However, the rank of $\varphi$ is $0\leq\varphi<2\pi$ which
means that there are identified points.  Note that the same
conclusions can be obtained if one starts with the Kruskal metric
(\ref{ds/krus}), absorbing $r_+$ in the coordinates $y^\alpha$, and
then letting $r_+\rightarrow 0$.

\subsection{Euclidean black hole and thermal anti-de Sitter space}
\label{Euc/4} 

The black hole constructed above has an Euclidean sector which can be
obtained from Euclidean anti-de Sitter space with identified points.
Indeed, consider the surface
\begin{equation}
x_0^2 + x_1^2 + x_2^2 + x_3^2 - x_4^2 = -l^2 \,,
\end{equation}
where we identify points along the boost in the plane $x_3/x_4$.
Following the same steps as before one obtains the Euclidean Kruskal
metric,
\begin{equation} 
ds^2 =  \frac{l^2(r+r_+)^2}{r_+^2}\, (dy_0^2 + dy_1^2 +dy_2^2)  + r^2
d\varphi^2, 
\label{ds/krusE}
\end{equation}
with \begin{equation}
r= r_+ \frac{1+y^2}{1-y^2} \,.
\end{equation}
The ranges are $-1\leq y_0,y_1,y_2 \leq 1$ with $0\leq y^2 <1$.  After
the identifications are done, the coordinate $\varphi$ has the range
$0\leq \varphi< 2\pi$.  The full Euclidean manifold is thus mapped
into a three dimensional solid ball $(0 \leq y^2< 1)$ times $S_1$.
The boundary is then the 2-sphere $y^2=1$ times $S_1$. 

The metric (\ref{ds/krusE}) can be put in a more familiar form by
making  the change of coordinates $\{\varphi,y^0,y^1,y^2\} \rightarrow
\{\varphi,\tilde r,\theta,\xi\}$:
\begin{eqnarray}
y^0 &=& f \sin\theta \sin\xi   \\
y^1 &=& f \sin\theta \cos\xi    \\
y^2 &=& f \cos\theta \,,
\end{eqnarray}
with $f = \tilde r/(l^2 + \sqrt{l^2 + \tilde r^2})$. One obtains
\begin{equation}
ds^2 = N^2 \left( r_+ d\varphi \right)^2 + 
\frac{d\tilde r^2}{N^{2}} + \tilde r^2 (d\theta^2 + \sin^2\theta
d\xi^2 )\,,
\label{u}
\end{equation}
with $N^2 = 1 + \tilde r^2/l^2$.  This metric clearly represents
Euclidean anti-de Sitter space with a ``time" coordinate $t =
r_+\varphi $. Actually, since $\varphi$ is compact, $0\leq \varphi
<2\pi$, (\ref{u}) should be called thermal anti-de Sitter space. The metric (\ref{u}) is the four dimensional version of the manifold considered in \cite{Witten98}.   The fact that we have end up with identified adS is not surprising since this was our starting point. 

The metric (\ref{u}) represents the Euclidean sector of our $M^3\times S_1$ black hole. We can now look at the black hole from a different point of view. Suppose we start with Euclidean anti-de Sitter space with metric (\ref{u}). Now we ask what is the hyperbolic, or Minkowskian sector, associated to this line element. The most natural continuation to Minkowski space is obtained by setting $\varphi \rightarrow i\varphi$ and yields Minkowskian anti-de Sitter space\footnote{Note that after continuing $\varphi$ to imaginary values one is forced to unwrap it in order to eliminate the closed timelike curves.}.   However, our previous discussion shows that one could keep $\varphi$ real and instead make the transformation $\xi\rightarrow i\xi$. This leads to an incomplete spacetime whose maximal extension give rise to the three dimensional Kruskal diagram shown before.  The idea of using the azimuthal coordinate as time was first discussed in \cite{BTZu}.  

The relation between thermal anti-de Sitter and the black hole is 
already present in the three dimensional situation. Indeed, consider
the spinless three dimensional black hole 
\begin{equation}
ds^2(r_+) = (-r_+^2 + r^2) dt^2 + \frac{dr^2}{-r_+^2 + r^2} + r^2
d\varphi^2 ,
\label{BTZ}
\end{equation}
with $r_+\leq r <\infty$, $0\leq t < \beta$ and $0\leq \varphi <2\pi$.
The period of $t$ is fixed by demanding the absence of conical
singularities in the $r/t$ plane and gives $\beta = 2\pi/r_+$. We
shall now prove that there exists a global change of coordinates that
maps (\ref{BTZ}) into the three dimensional thermal adS space. 

First define $r=r_+ \cosh\rho$. This maps (\ref{BTZ}) into
\begin{equation}
ds_1^2 = r_+^2 \sinh^2\rho dt^2 + d\rho^2 + r_+^2\cosh^2\rho
d\varphi^2 .
\label{3}
\end{equation}
Since $0\leq t <2\pi/r_+$, we can define $\theta = r_+ t$ which has
period $0\leq \theta <2\pi$. We also define $t' = r_+ \varphi$ which
has the period $2\pi r_+$. Finally we define $r' = \sinh\rho$
obtaining
\begin{equation}
ds_2^2(\beta') = (1 + r'^2) dt'^2 + \frac{dr'^2}{1 + r'^2} + r'^2
d\theta^2 ,
\label{TadS}
\end{equation}
with $0\leq r' <\infty$, $0\leq t' < 2\pi r_+$ and $0\leq \theta
<2\pi$.
The key element in this transformation is the permutation between the
angular and time coordinates. It should be evident that for the
Minkowskian black hole this transformation is not possible.  The
permutation between $t$ and $\varphi$ can be interpreted as a modular
transformation that maps the boundary condition $A_0=\tau A_\varphi$
into $A_0 = (-1/\tau) A_\varphi$ \cite{bbo}. It is then not a surprise
that the black hole had a temperature $\beta=2\pi/r_+$, while
(\ref{TadS}) has a temperature $\beta' = 2\pi r_+$. Indeed, the
corresponding modular parameters are related by a modular
transformation \cite{bbo}.  The line elements (\ref{BTZ}) and
(\ref{TadS}) have recently been investigated in the context of string
theory in
\cite{M-S}.

\section{The five dimensional topological black hole} \label{fivebh}

We have seen in the last section that by identifying points in anti-de
Sitter space one can produce a black hole with many of the properties
of the 2+1 black hole. The topology of the causal structure is however
different. In this section we shall describe how this procedure can be
repeated in five dimensions. Actually, the black hole exists for any
dimension \cite{b3} but the five dimensional case has some
peculiarities that makes it worth of a separate study, specially in
the Euclidean sector. 

\subsection{The spinless 5D black hole}

We shall not repeat here the geometrical construction of the 5D black
hole (it can be found in \cite{b3}). We only quote here the form of
the Kruskal metric which is a natural generalization of
(\ref{ds/krus}),
\begin{equation} 
ds^2 =  \frac{l^2(r+r_+)^2}{r_+^2}\, dy^\alpha
dy^\beta\eta_{\alpha\beta} + r^2 d\varphi^2, 
\label{ds/krus5}
\end{equation}
where $\alpha=0,1,2,3$, $-\infty < y^\alpha < \infty$ with $-1<y^2<1$,
$y^2=-y_0^2+y_1^2+y_2^2+y_3^2$ and $0\leq \varphi<2\pi$. The radial
coordinate $r$ depends on $y^2$
and it is given by
\begin{equation}
r = r_+ \frac{1+y^2}{1-y^2}.
\end{equation}
This metric represents the non-rotating black hole in Kruskal
coordinates. It is clear that the causal structure associated to this
black hole is four dimensional. The horizon is located at the three
dimensional hypercone $y_0^2=y_1^2+y_2^2+y_3^2$ while the singularity
at the three dimensional hyperboloid $y_0^2=1+y_1^2+y_2^2+y_3^2$. 

The Euclidean black hole can be obtained simply by setting $y_0
\rightarrow i y_0$. The Euclidean metric can be put in the familiar
spherically symmetric form by introducing polar coordinates in the
hyperplane $y_\alpha$, 
\begin{equation}
ds^2 = N^2 (  r_+ d\varphi)^2 + \frac{d\tilde r^2}{N^2} + \tilde r^2
d\Omega^2_3 \,,
\label{Eads/5}
\end{equation}
with $0\leq \varphi <2 \pi$. As in four dimensions this metric
represents Euclidean anti-de Sitter with identified points, or 5d
thermal anti-de Sitter space, if the coordinate $\varphi$ is treated
as
time. We shall indeed use this interpretation in the next section to
define global charges.  

Just as in the four dimensional case, the metric can be put in a
Schwarzschild form and, in particular, in the Euclidean sector these
coordinates are complete.  The metric is given by
\begin{equation}
ds^2 = l^2 N^2 d\Omega_3 +  N^{-2} dr^2 + r^2 d\varphi^2, 
\label{ds/sch5}
\end{equation}
with $N^2(r) = (r^2 -r_+^2)/l^2 $, and
\begin{equation}
d\Omega_3 = -\sin^2{\theta}\, dt^2 + \frac{l^2}{r_+^2} (d\theta^2 +
\sin^2{\theta} d\chi^2) .
\label{sph3}
\end{equation}
The ranges are $0 < \theta< \pi/2$, $0\leq\chi<2\pi$, $0\leq\varphi <
2\pi$ and $r_+<r<\infty$. 

\subsection{Euclidean anti-de Sitter space} 

Our aim now is to construct a rotating solution. We shall start with
the Euclidean case because in that sector we can prove explicitly
that the resulting manifold is complete and corresponds to Euclidean
anti-de Sitter space with identified points. The continuation to
Minkowskian signature will be studied below. 

We start considering Euclidean anti-de Sitter space. In five dimensions, 
this space has three commuting Killing vectors. In this section we shall write the induced metric in a set of coordinates for which those isometries are manifiest.

Euclidean anti-de Sitter space in five dimensions is defined by
\begin{equation}
x_0^2 + x_1^2 + x_2^2 + x_3^2 + x_4^2 - x_5^2 =-l^2.
\label{Eads}
\end{equation}
This surface has two disconnected branches $x_5>l$ and $x_5<-l$ and we
take, for definiteness, $x_5>l$. We introduce polar coordinates in the
planes $x_0/x_1$, $x_2/x_3$ and $x_5/x_6$:
\begin{eqnarray}
x_0 = \lambda \sin\tilde t \ \ \ \  x_2 = \mu \sin\tilde \chi \ \ \ \ 
x_4 = \sigma \sinh\tilde \varphi \\
x_1 = \lambda \cos\tilde t \ \ \ \  x_3 = \mu \cos\tilde \chi \ \ \ \
x_5 = \sigma \cosh\tilde \varphi ,
\end{eqnarray}
which cover the full branch $x_5>l$. The radii $\lambda$, $\mu$ and
$\sigma$ are constrained by $\lambda^2 + \mu^2 -\sigma^2=-l^2$.
We now introduce polar coordinates in the plane $\lambda/\mu$,
\begin{equation}
\lambda = \nu \sin\theta, \ \ \ \ \mu = \nu \cos\theta.
\end{equation}
Note, however, that since $\lambda$ and $\mu$ are positive, the angle
$\theta$ has the rank $0\leq \theta <\pi/2$.  (In \cite{b3} the
-incorrect- rank $0\leq \theta <\pi$ was used.) Finally, $\nu$ and 
$\sigma$ are constraint to $\nu^2-\sigma^2=-l^2$, we thus introduce a
coordinate $\rho$ , 
\begin{equation}
\nu = l\sinh\rho, \ \ \ \  \sigma = l\cosh\rho.
\end{equation}
The induced metric can be calculated directly and one obtains
\begin{eqnarray}
\frac{ds^2}{l^2} &=& \sinh^2\rho [\cos^2\theta d\tilde t^2 +
\sin^2\theta d\tilde\chi^2 + d\theta^2]  \nonumber \\
&+& d\rho^2 + \cosh^2 \rho d\tilde \varphi^2
\label{E}
\end{eqnarray}
which, with the ranges $-\infty<\tilde\varphi<\infty$, $0\leq\tilde
t,\tilde \chi < 2\pi$, $0\leq\theta<\pi/2$ and $0\leq\rho<\infty$, 
cover the full branch $x_5>l$ of Euclidean anti-de Sitter space.
This metric will be our starting point in the construction of the
rotating black hole.

\subsection{A black hole with two times}

The key step in producing the black hole, in both the Euclidean and
Minkowskian signatures, are the identifications. We shall now make
identifications on the metric (\ref{E}). Before we redefine the
coordinates $\tilde t$ and $\tilde \chi$ as 
\begin{eqnarray}
\tilde t &=& A t + (b/l) \varphi \nonumber\\
\tilde \chi &=& A \chi + (b/l) \varphi \nonumber \\
\tilde\varphi &=& (r_+/l) \varphi \nonumber
\end{eqnarray}
with $A=(b^2 + r_+^2)/(l^2 r_+)$, where $r_+$ and $b$ are arbitrary
constants with dimensions of length. The metric becomes
\begin{eqnarray}
ds^2 &=& \cos^2\theta [l^2N^2 dt^2 + r^2(d\varphi + N^\varphi dt)^2] +
          \nonumber\\
     &+&  \sin^2\theta [l^2 N^2 d\chi^2 + r^2(d\varphi + N^\varphi
     d\chi)^2] \nonumber \\
 &+& N^{-2} dr^2 + l^2\frac{r^2-r_+^2}{r_+^2 + b^2} d\theta^2,
 \label{bhxbh}
\end{eqnarray}
with
\begin{eqnarray}
N^2 &=& \frac{(r^2-r_+^2)(r^2+b^2)}{l^2r^2}, \\
N^\varphi &=& \frac{b}{r_+} - \frac{r_+ b}{r^2},
\end{eqnarray}
and the radial coordinate $r$ is related to $\rho$ by 
\begin{equation}
r^2 = r_+^2 \cosh^2 \rho + b^2 \sinh^2 \rho.
\end{equation}
We now identify the points $\varphi \sim \varphi + 2\pi n$ obtaining
the Euclidean rotating black hole in five dimensions. 

The metric (\ref{bhxbh}) has a remarkable structure.  The lapse $N^2$
and shift $N^\varphi$ functions are exactly the same as those of the
2+1 black hole. The section $\theta=0$ represents a rotating 2+1 black
hole in the three dimensional space $r,t,\varphi$, while  the section
$\theta=\pi/2$ represents a rotating 2+1 black hole in the space
$r,\chi,\varphi$.  There is also an explicit symmetry under the
interchange of $t \leftrightarrow \chi$.  This metric can only exists
in the Euclidean sector, or in a spacetime with two times. Indeed, if
one continues back to a Minkowskian time by $t\rightarrow it$, one also
needs to change $\chi\rightarrow i\chi$ in order to maintain the
metric real.

\subsection{Minkowskian rotating black hole}

The Euclidean black hole that we have produced in the last section
cannot be continued back to Minkowskian space. In order to produce a
black hole by identifications on Minkowskian anti-de Sitter space we
need to make identifications along a different Killing vector. Let us
go back to the metric (\ref{E}), make the replacement $\tilde t
\rightarrow i\tilde t$, and redefine the coordinates as 
\begin{eqnarray}
\tilde t &=& \left(\frac{r_+^2 - r_-^2}{l^2r_+} \right) t + \frac{r_-
\varphi}{l}  \nonumber\\ 
\tilde\varphi &=& \frac{r_+}{l} \varphi \nonumber \\
\tilde\chi &=& \chi
\label{tranfo}
\end{eqnarray}
with $r_+$ and $r_-$ two arbitrary real constants with dimensions of
length. This converts (\ref{Eads}) into
\begin{eqnarray}
ds^2 &=& \cos^2\theta [-N^2 l^2 dt^2 + r^2(d\varphi + N^\varphi dt)^2]
+N^{-2}
dr^2  \nonumber \\
 &+&l^2\frac{r^2-r_+^2}{r_+^2 -r_-^2}(d\theta^2 +\sin^2\theta
 d{\chi}^2) 
 \nonumber \\ &+&  \frac{r^2-r_-^2}{r_+^2 -r_-^2}r_+^2\sin^2\theta
 d\varphi^2   ,
\label{dsam}
\end{eqnarray}
with
\begin{eqnarray} \label{nnphi}
N^2 &=& \frac{(r^2-r_+^2)(r^2-r_-^2)}{l^2 r^2}, \nonumber\\
N^\varphi &=& -\frac{r_-}{r_+}\frac{(r^2-r_+^2)}{r^2} \,,
\end{eqnarray}
and 
\begin{equation}
r^2 = r_+^2 \cosh^2 \rho -r_-^2 \sinh^2 \rho \,.
\end{equation}
Now we identify points along the coordinate $\varphi$:
\begin{equation}
\varphi \sim \varphi + 2\pi n, \ \ \ \ n \in Z.
\label{id}
\end{equation}
The metric (\ref{dsam}) with the identifications (\ref{id}) defines
the five dimensional Minkowskian rotational black hole.  An important
issue, that we shall not attack here, is to find the maximal extension
associated to this geometry. One can imagine the form of the Penrose
diagram  in the rotating case as a cylinder containing an infinite
sequence of alternating asymptotic and singularity regions.  However,
the explicit construction of the maximal extension has escape us. 
This problem is of importance because, as it has been pointed out in
\cite{Holst-Peldan}, the topological black hole is not static and
therefore the metric (\ref{dsam}) cannot be complete.  Indeed, it is
easy to prove that the replacement $\tilde t \rightarrow i\tilde t$ in
(\ref{E}) produces an hyperbolic metric with does not cover the
full Minkowskian anti-de Sitter surface. Note finally that the section
$\theta=0$ corresponds exactly to a rotating 2+1 black hole.  The
metric (\ref{dsam}) has a horizon located at $r=r_+$.

\subsection{The extreme and zero mass black hole} 

To find a extreme black hole from the general rotating case
(\ref{dsam}),
we define the non-periodic coordinate $\sigma$ by the relation
$l \theta=\sigma\sqrt{r_+^2-r_-^2} $, and take the limit $r_+ ^2
\rightarrow
r_-^2=a^2$.  The resulting metric is
\begin{eqnarray} 
\label{dsext}
ds^2 &=& \mp 2\frac{r^2-a^2}{l}d t d\varphi+\frac{l^2 r^2
dr^2}{(r^2-a^2)^2} \nonumber\\ &+& \left( r^2+
a^2\frac{r^2-a^2}{l^2}\sigma^2\right) d\varphi^2 \nonumber \\
&+&(r^2-a^2)(d \sigma^2 +\sigma^2 d\chi^2 ).
\end{eqnarray} 
The $\mp$ sign arises from the quotient $r_-/r_+$ in $N^\varphi$(see
(\ref{nnphi}). This metric represents the extreme black hole. It has
only one charge and one horizon and it also obtainable from anti-de
Sitter space by identifications. However, as in 2+1 dimensions, the
Killing vector needed is not the same as in the non-extreme case. We
shall explicitly prove this below and when we compute the group
element $g$ associated to the extreme black hole. The zero mass black
hole can be obtained from (\ref{dsext}) by letting $a=0$ and $\varphi\rightarrow \varphi \pm t$.

\section{Euclidean Global charges in five dimensions}

The issue of global charges requires to fix the asymptotic conditions
and the class of metrics that will be considered in the action
principle. 
We shall make a slight generalization of
the metric (\ref{dsam}) by considering again Euclidean anti-de Sitter
space with the metric (\ref{E}) and making the change of
coordinates
\begin{eqnarray}
      \tilde t &=& A t + b \varphi, \nonumber\\
\tilde \varphi &=& B t + r_+ \varphi,
\label{trans1}
\end{eqnarray}
with 
\begin{eqnarray}
A &=& r_+ \beta + b \Omega ,\nonumber\\
B &=& -b \beta + r_+ \Omega .
\label{AB}
\end{eqnarray}
This class of metrics contain the black hole in the case $B=0$ 
which implies $\Omega = \beta b/r_+$, $A= \beta(r_+^2 + b^2)/r_+$ and
(\ref{trans1}) reduces to (\ref{tranfo}).  The parameter $\beta$ allows 
us to fix $0<t<1$. For the black hole $\beta$ is fixed by the demand
of no conical singularities at the horizon as $\beta =
r_+/(b^2+r_+^2)$.
We consider here, however, only the asymptotic metric following from
(\ref{trans1}) with  $\beta$ and $\Omega$ fixed but arbitrary. 

A convenient choice for the tetrad is
\begin{eqnarray}
e^1 &=& l\sinh\rho \cos\theta (A dt+ bd\varphi) \nonumber\\
e^2 &=&l\sinh\rho \sin\theta d\chi \nonumber\\
e^3 &=&l\sinh\rho d\theta \nonumber\\
e^4 &=& ld\rho\nonumber\\
e^5 &=& l\cosh\rho (Bdt + r_+ d\varphi).
\label{eee}
\end{eqnarray}
The corresponding non-zero components of the spin connection are
\begin{eqnarray}
\omega^{13} &=& -\sin\theta (Adt+ bd\varphi) \nonumber\\
\omega^{14} &=& \cosh\rho \cos\theta (Adt+bd\varphi) \nonumber\\
\omega^{45} &=& -\sinh\rho (Bdt + r_+d\varphi )\nonumber\\
\omega^{23} &=& \cos\theta d\chi \nonumber\\
\omega^{24} &=& \cosh\rho \sin\theta d\chi \nonumber\\
\omega^{34} &=& \cosh\rho d\theta .
\label{www}
\end{eqnarray}
With these formulas at hand we can now compute the value of the
different charges associated to this metric. 

\subsection{Global charges in Chern-Simons gravity}

We shall now consider the five dimensional topological black holes
embedded in a Chern-Simons formulation for gravity and supergravity in
five dimensions \cite{b3}. We shall consider first a $SO(4,2)$
Chern-Simons theory which represents the simplest formulation of
gravity in five dimensions as a Chern-Simons theory. As we shall see
all charges associated to the topological black hole vanish in this
theory. Indeed, there is a curious cancellation of the contributions
to
the charges coming from the Einstein-Hilbert and Gauss-Bonnet terms.

We then consider  a supergravity theory constructed as a Chern-Simons
theory for the supergroup $SU(2,2|N)$ \cite{Chamseddine} which is the
natural extension of the three dimensional supergravity theory
constructed in \cite{Achucarro}.  Remarkably, the black hole also
solves the equations of motion following from this action and as such
the mass and angular momentum are different from zero.        
 
In five dimensions, a Chern-Simons theory is defined by a Lie algebra
$[J_A,J_B]=f^C_{\ AB} J_C$ possessing and invariant fully symmetric
three rank tensor $g_{ABC}$. The Chern-Simons equations of motion then
read,
\begin{equation}
\frac{1}{2}g_{ABC} F^B \ww F^C = 0  
\label{CSE}
\end{equation}
where ${\sf F} = d{\sf A} + {\sf A} \ww {\sf A}$ is the Yang-Mills
curvature for the Lie algebra
valued connection ${\sf A}=A^A J_A$.   The solutions that we have
considered have ${\sf F}=0$ and thus ${\sf A}=g^{-1} dg$ where $g$ is
a map from the manifold to the gauge group. However, due the
identifications the black hole map  $g$ is not single valued. In other
words, if ${\sf A}$ represents the flat connection associated to the
black hole, then the path ordered integral along the non-trivial loop
$\gamma$ generated by the identifications $P
\exp{\oint_\gamma {\sf A}}$ is different from one. 

In three dimensions, where the equations are simply $g_{AB} F^B=0$, it
is now well established that the Hilbert space of a Chern-Simons
theory is described by a conformal field theory\cite{Witten89}.  For
manifolds with a boundary, the underlying conformal field theory is a
Chiral WZW model \cite{Moore} whose spectrum is generated by Kac-Moody
currents.  From the point of view of classical Chern-Simons theory,
these results can be derived by studying global charges \cite{Bal,B}. 

It is remarkable that part of the results valid on three dimensions
are carried over in the generalizations to higher odd-dimensional
spacetimes. Indeed, it has been proved in \cite{BGH2} that in the
canonical realization of the gauge symmetries the constraints satisfy
the algebra of the $WZW_4$ theory found in \cite{Nair}. The $WZW_4$
theory is a natural generalization to four dimensions of the usual
$WZW$ theory.  A key property of the $WZW_4$ theory is the need of a
K\"ahler form (an Abelian closed 2-form). This two form, which greatly
facilitates the issue of global charges, appears naturally in five
dimensional supergravity\cite{Chamseddine,B-Tr-Z}. 

Global charges can easily be constructed from the Chern-Simons
equations of motion (\ref{CSE}). Let $E_A$ be the Chern-Simons
equations of motion and let $\delta A^A$ be a perturbation of the
gauge field not necessarily satisfying the linearized equations. Using
the Bianchi identity $\nabla F^A=0$ and the invariant property of
$g_{ABC}$ ($\nabla g_{ABC}=0$) it is direct to see that $\delta E_A$
is given by,
\begin{equation}
\delta E_A= \nabla(g_{ABC} F^B \ww \delta A^C).
\end{equation}
Now, let $\lambda^A$ a Killing vector of the background configuration
($\nabla \lambda^A=0$), then the combination $\lambda^A \delta E_A$ is
a total derivative,
\begin{equation}
\lambda^A \delta E_A = d(g_{ABC} \lambda^A F^B \ww \delta A^C),
\end{equation}
and thus conserved $d(\lambda^A \delta E_A)=0 $. Hence, for every
Killing vector $\lambda^A$ there is one conserved current. We now
consider a manifold with the topology $\Sigma\times \Re$ and we assume
that $\Sigma$ has a boundary denote by $\partial \Sigma$. The integral
$\int_\Sigma \lambda^A \delta E_a$ is thus independent of $\Sigma$ and
provides a charge at $\pp \Sigma$ equal to
\begin{equation}
\delta Q(\lambda) = \int_{\pp \Sigma} g_{ABC} \lambda^A F^B \ww \delta
A^C.
\label{dQ}
\end{equation}
This formula gives the value for the variation of $Q$. The problem now
is to extract the value of $Q$ from (\ref{dQ}). For this the specific
form of the boundary conditions is necessary.

\subsection{The $SO(4,2)$ theory. Zero charges. }

Consider now the case on which the Lie algebra is $SO(4,2)$ generated
by $J_{AB}$. Note the change in notation, each pair $(A,B)$ correspond
to $A$ in (\ref{CSE}).  This algebra indeed has a fully-symmetric
invariant three rank tensor, namely, the Levi-Cevita form
$\epsilon_{ABCDEF}$. The equations of motion read
\begin{equation}
\epsilon_{ABCDEF} \tilde R^{AB} \wedge \tilde R^{CD} =0 .
\label{csgr}
\end{equation}
The link with general relativity is achieved when one defines the
$\omega^{a6}$ component of the gauge field as the vielbein: $e^a =
l\omega^{a6}$. (The arbitrary parameter $l$, with dimensions of
length, is
introduced here to make $e^a$ a dimensionful field and will be related
to the cosmological constant.)  Once this identification is done, the
component $R^{a6}$ of the curvature becomes equal to the torsion:
$T^a= l\tilde R^{a6} = De^a$ with $D$ the covariant derivative in the
spin connection $\omega^{ab}$. 

The equations (\ref{csgr}) can be shown to come from a Lagrangian
containing a negative cosmological constant $(-1/l^2)$, the
Einstein-Hilbert term, and a Gauss-Bonnet term (which in five
dimensions is not a total derivative),
\begin{equation}
L = \sqrt{-g} [k {\bf R}^2 + (1/G) R + \Lambda],
\label{L}
\end{equation}
where ${\bf R}^2$ denotes the Gauss-Bonnet combination. The couplings
$k,\Lambda$ and $G$ are not arbitrarily but linked by the $SO(4,2)$
symmetry \cite{Chamseddine,BTZ2}. The topological black hole solves
the above equations simply because it has zero torsion and constant
spacetime curvature. These two conditions imply $\tilde R^{AB}=0$ and
thus (\ref{csgr}) is satisfied.     

Going back to our equation for the charges we can see that they vanish
for the black hole. Indeed, given a Killing vector $\eta_{AB}$ the
charge associated, up to constants, would be
\begin{equation}
\delta Q \sim \int \epsilon_{ABCDEF} \eta^{AB} \tilde R^{CD} \delta
\omega^{EF},
\end{equation}
but since for the black hole $\tilde R^{AB}=0$, $\delta Q=0$ and thus
the charge vanishes, up to an additive fixed constant.  

This result shows a curious cancellation in the value of the charges
associated to the Einstein-Hilbert and Gauss-Bonnet terms. One can
prove that the charges associated to the theory containing only the
Hilbert term (plus $\Lambda$) are divergent quantities.  We have
just proved that the full Lagrangian has zero charges, this implies a
cancellation between the contributions to the global charges coming from  the first and last two terms in (\ref{L}). \\

\subsection{Charges in the $SO(4,2)\times U(1)$ theory. }

Now we consider the particular gauge group $SO(4,2)\times U(1)$. This
group arises in the Chern-Simons supergravity action in five
dimensions. The orthogonal part $SO(4,2)$ is just the anti-de Sitter
group in five dimensions while the central piece, $U(1)$, is necessary
to achieve supersymmetry\cite{Chamseddine,B-Tr-Z,Tr-Z}. It is
remarkable that the factor $U(1)$ provides a great simplification in
the analysis of the canonical structure as well as the issue of global
charges. The coupling between the geometrical variables and the $U(1)$
field, denoted by $b$, is simply $\tilde R^{AB} \wedge \tilde R_{AB}
\wedge b$. The equations of motion are modified in this case to 
\begin{eqnarray}
\epsilon_{ABCDEF} 
\tilde R^{AB} \wedge \tilde R^{CD} &=& \tilde R_{EF} \wedge K ,\\
\tilde R^{AB} \wedge \tilde R_{AB} &=& 0, 
\end{eqnarray}
where $K=db$. As before, we identify $e^a =l\omega^{a6}$ which implies
$T^a=l\tilde R^{a6}=De^a$.  Note that the topological black hole also
solves the above equations of motion because they have constant
curvature ($\tilde R^{ab} =0$) and zero torsion ($T^a=0$). $K$ is left
arbitrary.    

Applying the formula (\ref{dQ}) to this set of equations of motion,
using the boundary conditions $\tilde R^{AB}=0$, which are satisfied
by the black hole, and fixing the value of $b$ (and hence $K=db$), we
obtain the value of $\delta Q$,
\begin{equation}
\delta Q = \int_{\pp \Sigma} K \ww \delta \omega^{AB} \lambda_{AB} .
\label{Qcs}
\end{equation}
where $\lambda_{AB}$ is a Killing vector of the background
configuration $\nabla \eta^{AB}=0$. 

Note that $K$ is left arbitrary at the boundary (it appears in the
equations of motion multiplied by $\tilde R^{AB}$). Hence it is
consistent with the dynamics to fix its value at the boundary.  In
terms of $e^a$ and $\omega^{ab}$ this formula reads, 
\begin{equation}
\delta Q[\eta_a,\eta_{ab}] = \int_{\partial \Sigma} K
\wedge (2  \delta e^a \eta_a- l \delta \omega^{ab}\eta_{ab}).
\label{Q}
\end{equation}
This formula for the global charge depends crucially on the assumption
that the topology of the manifold is $\Sigma \times \Re$ (or $\Sigma
\times S_1$)  and that $\Sigma$ has a boundary.  The black holes that
we have been discussing do have this topology, however, the $\Re$
($S_1$) factor does not represent time, but rather, the angular
coordinate. Indeed, the five dimensional black hole has the topology
$B_4 \times S_1$ where $B_4$ is a 4-ball whose boundary is
$S_3$\footnote{In \cite{b3} $\Sigma$ was taken to be $S_2\times S_1$.
This topology arises from an error in the rank of the coordinate
$\theta$ whose right value is $0\leq\theta <\pi/2$, as opposed to
$0\leq \theta <\pi$ as in \cite{b3}. Surprisingly, the actual value of
the charges is insensitive to this problem and the same value of $M$
and $J$ are obtained here. See Eq. (\ref{dqf}) below.}. As discussed
before, the Euclidean black hole can be interpreted as thermal anti-de
Sitter space provided one treats the angular coordinate as time. We
shall now compute global charges for the black hole treating $\varphi$
as the time coordinate.  In the formula (\ref{Q}) $\partial\Sigma$
thus represents the three-sphere parameterized by the coordinates
$t,\chi,\theta$. 

The formula (\ref{Q}) depends on the 2-form $K$ which was not
determined by the equations of motion. The only local conditions over
$K$ are $dK=0$ (since $K=db$), and that it must have maximum rank
\cite{BGH2}. In particular, $K$ must be different from zero
everywhere.  It turns out that global considerations suggest a natural
choice for the pull back of $K$ into $\partial \Sigma=S_3$. Since $K$
is a two form, its dual is a vector field. We then face the problem of
defining a vector field on $S_3$, different from zero everywhere. As
it is well known there are not too many possibilities. In the
coordinates $t,\chi,\theta$ that parameterized the sphere in the
metric (\ref{E}) with the change (\ref{trans1}) two natural
choices for the dual of $K$ are: $^*K_t = \partial_t$ or $^*K_\chi= 
\partial_\chi$.  It is direct to see that $^*K_\chi $ does not give
rise to any conserved charges. We shall  then take $^* K
=(k/\pi^2)\partial_t$ with $k$ an arbitrary but fixed constant. [The
normalization factor
is included for convenience $\pi^2 = \int d\theta d\chi$.]      

The formula for the charge thus becomes
\begin{equation}
\delta Q[\eta_a,\eta_{ab}] = \frac{k}{\pi^2} \int_{S_3}
( 2 \eta_a \delta e^a_t - l\eta_{ab} \delta \omega^{ab}_t)dS . 
\label{Q2}
\end{equation}

The black hole geometries described above have three commuting Killing
vectors, $\partial_t$, $\partial_\chi$ and $\partial_\varphi$. These
metric Killing vectors translate into connection Killing vectors via 
$\xi^\mu A_\mu$. Thus, for example, invariance under translations in
$\varphi$ is reproduce in the connection representations as invariance
of the connection under gauge transformations with the parameter
$\{e^a_\varphi, \omega^{ab}_\varphi\}$. From the formulas (\ref{eee})
and
(\ref{www}) one can compute the charge associated to this invariance
and obtain
\begin{equation}
\delta Q[e^a_\varphi, \omega^{ab}_\varphi] = \beta \delta (2k br_+) +
\Omega \delta(k(r_+^2 - b^2)).
\label{dqf}
\end{equation}
The interpretation for (\ref{dqf}) is straightforward: $\beta$ and
$\Omega$ are the conjugates of $M= 2kb r_+$ and $J = k(r_+^2 -
b^2)$.  

Note that $k$, which acts as a coupling constant, is not an universal
parameter in the action but the -fixed- value of the $U(1)$ field
strength $K$. This is similar to what happens in string theory where
the coupling constant is equal to the value of the dilaton field at
infinity.

We have just shown that the parameters $r_+$ and $b$ give rise to
conserved
global charges. In the next section we shall compute the holonomies
existing in the black hole geometry and  show that $\{r_+,b\}$ also
represent gauge invariant quantities.

\section{Group element. Holonomies}   

Since the topological black holes have constant spacetime curvature,
their anti-de Sitter Yang-Mills curvature is equal to zero. Thus, they
can be represented as a mapping $g$ from the manifold to the gauge
group $G$. This mapping is not trivial due to the identifications
needed to produce the black hole. Here we shall exhibit the explicit
form of the map $g$ and extract relevant information from it such as
the value of the holonomies, the gauge invariant quantities, the
possibility of finding Killing spinors, and the temperature of the
black hole. In three dimensions the group element was calculated in
\cite{CLM}.

A word of caution concerning global issues is necessary.
As we have pointed out above, the rotating black hole is known only in
the Schwarzschild coordinates which, in Minkowskian signature, do not
cover the full exterior manifold. This problem is solved when $J=0$
passing to Kruskal coordinates.  Here we shall be interested in the
holonomies generated by going around the angular coordinate $\varphi$,
and for this purpose the Schwarzschild coordinates are good enough.   

We shall make use of the spinorial representation of the so(4,2)
algebra,
given by the matrices $ J_{a b}=\Gamma_{a b}= \frac{1}{4}[\Gamma_{a},
\Gamma_{b}]$ and $J_{a,6}=\frac{1}{2}\Gamma_{a}$, where $\Gamma_{a}$
are
the Dirac matrices in five dimensions. Here $A=\{a,6\}$ with
$a=1,\cdots,5$.  We use the following representation of Dirac
matrices:
$$
\Gamma_m=i \left[ \begin{array}{cccc}
  0  &  -\sigma_{m}  \\
  \bar\sigma_{m}  &0
  \end{array}\right] \, ,
$$
with $\sigma_{m}=(-1,{\bf \sigma})$, $\bar{\sigma}=(1,{\bf
\sigma})$, $m=1,\cdots,4$ and
$\Gamma_5=\imath \Gamma_1 \Gamma_2 \Gamma_3 \Gamma_4 $.  For later
reference we remind that in five dimensions there exists two
irreducible
independent representations for the Dirac matrices, namely $\Gamma_m$
and
$-\Gamma_m$.   

The gauge field {\sf A} in the spinorial representation in terms of
vielbein $e^a$ and spin connection $\omega^{ab}$ is defined as, 
\begin{equation}
{\sf A}=\frac{e^a}{2l} \Gamma_a+\frac{1}{2}\omega^{ab}\Gamma_{ab}.
\label{As} 
\end{equation}
For notational simplicity, in this section we will set $l=1$.

\subsection{Group element of 5D Minkowskian rotating black hole}
\label{secxxx1}

To compute $g$ for the rotating black hole it is convenient to begin
with the group element for the sector of anti-de Sitter space obtained
from (\ref{E}) with the change $t \rightarrow it$. The vielbein
and spin connection for this metric are given in that section.  

The group element $g$ is related to gauge field {\sf A} by ${\sf
A}=g^{-1} dg$ and {\sf A} is given in terms of $e$ and $\omega$ in
(\ref{As}). The equations that determines $g$ thus are
\begin{eqnarray}
\partial_{\tilde t} g &=& g (\frac{1}{2}\sinh \rho \cos \theta
\Gamma_1+
 \cosh \rho \cos \theta \Gamma_{12}- \sin \theta \Gamma_{14}) \,,
 \label{eqt} \nonumber\\
\partial_\rho g &=&\frac{1}{2 } g  \Gamma_2\,, \label{eqrho}
\nonumber\\
\partial_{\tilde \varphi} g &=& g (\frac{1}{2} \cosh\rho
\Gamma_3-\sinh\rho
\Gamma_{23})\,, \label{eqphi} \nonumber\\
\partial_\theta g &=& g( \frac{1}{2}\sinh \rho
\Gamma_4 -\cosh \rho \Gamma_{24})\,,\label{eqtheta} \nonumber \\
\partial_\chi g &=& g (\frac{1}{2}\sinh\rho \sin \theta \Gamma_5
- \cosh \rho \sin \theta 
\Gamma_{25} -\cos \theta \Gamma_{45})\,,\label{eqchi} \nonumber
\end{eqnarray}     
whose solution is 
\begin{equation}  \label{gestam}
\displaystyle{ g_{adS}=g_0 e^{\frac{1}{2} \Gamma_3 {\tilde \varphi}}
e^{\Gamma_{12} {\tilde t}}
e^{\Gamma_{54}\chi}e^{\Gamma_{42}\theta}e^{\frac{1}{2 l} \Gamma_2
\rho}} ,\nonumber
\end{equation}
with $g_0$ an arbitrary constant group element. 

The black hole metric is obtained from (\ref{E}) by making the
transformation (\ref{tranfo}) and identifying points along the angular
coordinate $\varphi$. Thus, we perform the same coordinate
transformation in 
(\ref{gestam})  obtaining the group element for the rotating
black hole
\begin{equation}  \label{grot}
\displaystyle{ g_{BH}=g_0 e^{ p_{\varphi} \varphi}
e^{p_t  t} e^{-\Gamma_{45}\chi}e^{-\Gamma_{24}\theta}e^{\frac{1}{2 }
 \Gamma_2 \rho}} ,
\end{equation}
where $p_\varphi=r_+\Gamma_3/2+r_-\Gamma_{12}$ and 
$p_t=\frac{r_+^2 - r_-^2}{r_+}\Gamma_{12}$. 

The group element $g$ encodes gauge invariant information through
its holonomies.  First note that 
\begin{equation}
h := g_{\varphi=2\pi} g^{-1}_{\varphi=0} = e^{2\pi
(r_+\Gamma_3/2+r_-\Gamma_{12})} \neq 1.
\end{equation}
We now define $h=e^W$ and compute the Casimirs associated to $W=W^{AB}
J_{AB}$,  
\begin{eqnarray}
{\rm Tr}\,W^2 &=& 4 \pi^2 ( r_-^2+r_+^2 ) \,, \\
{\rm Tr}\,W^4 &=& 4 \pi^4 ( r_-^4+6 r_-^2 r_+^2+r_+^4 ).
\end{eqnarray}
Odd powers of $W$ vanish because $W^{AB}$ is antisymmetric.  The fact
that $r_+$ and $r_-$ are related to the Casimirs of the holonomy
confirm that they are gauge invariant quantities. 

The only remaining independent Casimir allowed by the Cayley--Hamilton
theorem is 
$$C_3=\epsilon_{ABCDEF} W^{AB} W^{CD} W^{EF},$$
which is zero
for the black hole with two charges.  In order to have $C_3\neq 0$ we
would need to consider identifications along a Killing vector with
three parameters of the form $\xi = r_1 J_{01} + r_2 J_{23} + r_3
J_{45}$. Note that $J_{01}, J_{23}$ and $J_{45}$ are commuting and
therefore $r_1,r_2$ and $r_3$ would represent observables of the
associated spacetime. It is actually not difficult to produce a metric
with three charges, but for our purposes here it is of no relevance so
we skip it.

\subsection{Group element for the extreme  black hole}

As we have seen in previous sections, the extreme metric cannot be
obtained in a smooth way from the non-extreme one. It should then be
not a surprise  that the associated group elements cannot be deformed
one into the other. This fact reflects a deeper issue:  The Killing
vectors that produce the non-extreme and extreme metrics belongs to
different classes and one cannot relate them by conjugation by an
element of the anti-de Sitter group. This was already the case in
three dimensions  \cite{BHTZ}, the novelty here is that, contrary to
the 2+1 case, the metrics cannot be deformed one into the other.    

The metric for the extreme black hole was displayed in (\ref{dsext}).    
The  vielbein can be chosen as
\begin{eqnarray}
e^1 &=& -\frac{e^{ \rho}}{F} dt \nonumber\\
e^2 &=& d\rho  \nonumber\\
e^3 &=& \mp \frac{e^{ \rho}}{F} dt +F e^{ \rho} d\varphi \nonumber\\ 
e^4 &=& e^{\rho} d\sigma \nonumber\\
e^5 &=& e^{\rho} \sigma \, d\chi \,,
\label{vielbeinext2}
\end{eqnarray}
and the non zero components of the spin connection are
\begin{eqnarray}
\omega^1{\,}_2&=&-\frac{1}{F}(e^{\rho}\,dt\mp a^2 e^{- \rho}\,
d\varphi)  \\
\omega^1{\,}_3&=&\frac{a^2  }{F^2 } (\mp e^{-2 \rho}\,d\rho \pm \sigma
\,
d\sigma ) \nonumber\\
\omega^1{\,}_4&=&\pm \frac{a^2  \sigma }{F } d\varphi \nonumber\\
\omega^2{\,}_3&=&\frac{ e^{\rho}  }{F }(\pm dt-
   (1+a^2 \sigma^2) \,d\varphi ) \nonumber \\
\omega^2{\,}_4 &=& -e^{\rho} \, d\sigma \nonumber \\
\omega^2{\,}_5 &=& -e^{\rho} \sigma \, d\chi \nonumber\\
\omega^3{\,}_4 &=& \frac{a^2  \sigma }{ F} \,d\varphi \nonumber\\ 
\omega^4{\,}_5 &=& -d\chi,  
\label{spinconnectionext2}
\end{eqnarray}
with $F=\sqrt{1+a^2(\sigma^2+e^{-2\rho})}$ and $\rho$ is related to
radial
coordinate $r$ of (\ref{dsext}) by $\rho=\log(r^2-a^2)^{1/2}$ .

The differential equation for $g$ is now more complicated but it can
still be integrated. After a rather long  but direct calculation we
find the group element for the extreme black hole
\begin{equation}  \label{gext}
 \displaystyle{ g=g_0 e^{ \frac{1}{2}p_{\varphi} \varphi}
e^{p_t  t}
e^{-\Gamma_{45}\chi}e^{(\Gamma_4/2-\Gamma_{24})\sigma}e^{\frac{1}{2 }
 \Gamma_2 \rho \pm \log F \Gamma_{13}}} ,
\end{equation}
where $p_\varphi=\pm a^2 (
\Gamma_1/2-\Gamma_{12})+(a^2+2)\Gamma_3/2+(a^2-2)\Gamma_{23}$, 
$p_t=-\Gamma_1/2-\Gamma_{12} \mp (\Gamma_3/2-\Gamma_{23})$,
 and $g_0$ is a constant matrix.    

We note that this group element cannot be derived from the general
case (\ref{grot}) by taking the limit $r_+\rightarrow r_-$. As
stressed before,  this reveals that the Killing vectors used to
construct these black holes belong to a distinct classes in the
isometry group.   The value of the holonomy in this case is 
($h=e^{\pi W}$) 
\begin{equation}
W = \pm a^2
(\frac{\Gamma_1}{2}-\Gamma_{12})+(a^2+2)\frac{\Gamma_3}{2}
+(a^2-2)\Gamma_{23} 
\end{equation}
and the corresponding invariants are
\begin{equation}
{\rm Tr}\,W^2= 8 \pi^2  a^2 \quad \mbox{and} \quad
{\rm Tr}\,W^4= 32 \pi^4  a^4\,.
\end{equation}
  
The group element associated to the massless black hole can be
obtained by letting $a\rightarrow 0$ in the extreme $g$.   One finds 
\begin{equation}  \label{gm0}
 \displaystyle{ g=g_0 e^{ p_{\varphi} \varphi}
e^{p_t  t}
e^{-\Gamma_{45}\chi}e^{(\Gamma_4/2-\Gamma_{24})\sigma}e^{\frac{1}{2 }
 \Gamma_2 \rho }} ,
\end{equation}
where $p_\varphi=\Gamma{_3}/2-\Gamma_{23}$, 
$p_t=-\Gamma_1/2-\Gamma_{12} \mp (\Gamma_3/2-\Gamma_{23})$,
 and $g_0$ is a constant matrix.

\subsection{Group element of 5D Euclidean black hole}
\label{secgeucl}

We finally consider the group element associated to the Euclidean
black hole. The spinorial representation of the so(5,1) algebra can be
obtained from that of so(4,2) by changing $\Gamma_1$ by $i \Gamma_1$
(and therefore $\Gamma_5$ by $i \Gamma_5$). 

To calculate the group element for Euclidean black holes  we proceed
in the same way as in Minkowskian case. The analytic continuation does
not modifies the functional form of the group element for the portion
of anti-de Sitter space (\ref{E}). The starting point is thus its
group element (\ref{gestam}) written  in terms of
$\tilde{\tau}=i\tilde{ t}$ instead of  $\tilde{t}$ 
 \begin{equation}  \label{geuclads}
\displaystyle{ g=g_0 e^{\frac{1}{2} \Gamma_3 {\tilde \varphi}}
e^{\Gamma_{12} {\tilde \tau}}
e^{-\Gamma_{45}\chi}e^{-\Gamma_{24}\theta}e^{\frac{1}{2 } \Gamma_2
\rho}} .
\end{equation} 
 
The Euclidean rotating black hole is obtained making the
transformation
\begin{equation}
\tilde \tau = A \tau + b \varphi \,, \qquad  \tilde\varphi = r_+
\varphi \,,
\end{equation}
with $A=(b^2 + r_+^2)/r_+$, which applied on (\ref{geuclads}) yields
\begin{equation}  \label{geucrot1}
\displaystyle{ g = g_0 e^{(\frac{1}{2}r_+ \Gamma_3+b \Gamma_{12}) {
\varphi}}
e^{A\Gamma_{12}  \tau}
e^{-\Gamma_{45}\chi}e^{-\Gamma_{24}\theta}e^{\frac{1}{2 } \Gamma_2
\rho}}.
\end{equation} 

\subsection{Killing spinors}  \label{secKS}

An immediate application for the group elements calculated in the
previous sections is to analyse the issue of Killing spinors. Killing
spinors $\epsilon$ are defined  by 
\begin{equation}  \label{Killingeq}
{\sf D}\epsilon \equiv d\,\epsilon + {\sf A}\,\epsilon=0 \,,
\end{equation}
where $\sf D$ is the covariant derivative in the spinorial
representation. 
It is direct to see that $\epsilon= g^{-1} \epsilon_0$, with
$d\,\epsilon_0=0$ ( constant spinor), is the general solution of
(\ref{Killingeq}) provided that $g^{-1} dg={\sf A}$

As in 2+1 dimensions \cite{CH}, the five dimensional black hole has
locally as many Killing spinors as anti-de Sitter space, but only some
of them are compatible with the identifications. The global Killing
spinors will be
those (anti-) periodic in $\varphi$ or those independent of $\varphi$.
  
We begin with the extreme massive case.  Since that $g$ in
(\ref{gext}) is non-periodic in $\varphi$, it is necessary to choose
$\epsilon_0$ such that the angular dependence is eliminated. We find
that there is only one possibility for each sign in the limit $r_+
\rightarrow \pm r_-$, and $\epsilon_0$ must be
proportional to
$$ \left( \begin{array}{c}
i \\ \pm 1 \\ \mp i  \\ 1 
\end{array}\right) \,.$$

As mentioned before, there exists other inequivalent spinorial
representation of so(4,2).  We note that in this new representation
$p_\varphi$ does not change as a function of the Dirac matrices, but
its explicit representation is modified. We find that there is one
Killing spinor for each sign provided that $\epsilon_0$ is
proportional to 
$$ \left( \begin{array}{c}
-i \\ \mp 1 \\ \mp i  \\ 1 
\end{array}\right) \,.$$

The $\varphi$ dependence of $\epsilon$ in the $M=0$ black hole is
fixed by
\begin{equation}  \label{phidep}
\exp(-p_{\varphi} \varphi) =1-p_{\varphi}\varphi 
\end{equation}
given that $p_{\varphi}=\frac{1}{2}\Gamma_3+\Gamma_{23}$ verifies
$p_{\varphi}^2=0$.
Clearly, the global Killing spinors occurs only if $\epsilon_0$ is a
null eigenvector of $p_{\varphi}$. As this matrix has two linearly
independent null eigenvector there are two global Killing spinors for
the massless black hole. The same occurs in the other spinorial
representation.

As in 2+1 dimensions, the vacuum solution $r_+=r_-=0$ has the maximum
number of supersymmetries: two for each representation of the $\gamma$
matrices.


\subsection{Temperature from the group element.}

As a final application of the group elements we calculate here the
temperature of the Euclidean black hole.  In the Chern-Simons
formulation  there is, in principle, no metric and one cannot impose
the usual non-conical singularity condition on the Euclidean spacetime
to determine the Euclidean period, or inverse temperature. However,
there exists an analog condition which can be implemented with the
knowledge of $g$ and give the right value for the temperature.   

We can compute the temperature by imposing that $g$ is single valued
as one turns in the time direction,  
\begin{equation}
g(\tau=\beta) g^{-1}(\tau=0)=1 \,.
\end{equation}
In the case of rotating black hole this condition reads
\begin{equation} \label{condT}
\exp\{ \frac{b^2 + r_+^2}{ l^2 r_+}\beta J_{12} \}=1 \,.
\end{equation}
Notice that we now use the vectorial representation $J_{12}$ instead
of the spinorial representation $\Gamma_{12}$,  and $l$ has been
reincorporated. Since $J_{12}$ is a rotational generator we find the
condition $\frac{b^2 + r_+^2}{ l^2 r_+}\beta =2\pi$ which gives the
right value for $\beta$ \cite{b3}.


\section{Other solutions and final remarks}

In the last sections we have discussed black holes with the topology
$M^{D-1} \times S_1$. These black holes have constant curvature and
are constructed by identifying points in anti-de Sitter space.  A
natural question now arises. Could we find other black hole solutions,
not necessarily of constant curvature, with topologies of the form
$M^{k} \times W_{D-k}$ with $W$ a compact manifold?

\subsection{The $M^k\times W_{D-k}$ black hole ansatz}

Consider a $D$--dimensional space--time with a metric of the form
\begin{equation}
ds^2 = H(s)\eta_{ab}dx^{a}dx^{b} + r^2(s)d\Omega^{2}_{D-k} \ \ ,
\label{kruskal}
\end{equation}
with $\eta_{ab}=\mbox{diag}(-1,1,\ldots,1)$ in the $k$ first
coordinates $x^{a}$, $s=\eta_{ab}x^{a}x^{b}$, and $d\Omega^2$ is the
metric of a $D-k$--dimensional compact Euclidean space ($W$), 
\begin{equation}
d\Omega^2 = g_{ij}(y)dy^{i}dy^{j} \ \ \ \ i,j=k+1,...,D.
\end{equation}

For $k =1$, the above metric represents a cosmological--like solution.
For $k=2$, the conformally flat $2$--dimensional space of coordinates
$x^a$ can be seen, in some cases, as a Kruskal diagram for a
$M^2\times S_{D-2}$ black hole. For a generic value of $k$ we would
have a $M^{k} \times W_{D-k}$ black hole.  However, for this
affirmation  to be true, the following properties must be satisfied: 
\begin{enumerate}
\item[(i)] The functions $H(s)$ and $r(s)$ must be well behaved at the
point
$s=0$, which will represent black hole horizon.

\item[(ii)] The function $r$  must be positive at the horizon, $r(s=0)
=: r_+>0$. For $s>0$, $r$ must be a crescent function and go to
infinite for $s\rightarrow s_{\infty}$, where  $s_{\infty}$ can be
either infinity or some positive finite number. For $s<0$ the function
$r$ must to go to zero at some point $s=s_{0}$ and the spacetime
geometry should be singular there.

\item[(iii)] The function $H$ must be positive for all values of $s$.

\item[(iv)] $W$ must be a compact space. 
\end{enumerate}

When these properties are fulfilled, for a given value of $k$, we will
say that the metric (\ref{kruskal}) represents a black hole with
topology  $M^{k} \times W_{D-k}$. Of course the Schwarzschild and
topological black holes satisfy (i)-(iv) with $k=2$ and $k=D-1$,
respectively.  The problem now is to find solutions to Einstein
equations (or its generalizations) satisfying (i)-(iv) with different
values of $k$, and different choices of $W$.     

We have investigated the conditions imposed by the standard Einstein
equations on the functions $H$ and $s$, with and without cosmological
constant. For $k>2$ the equations become difficult to treat and we did
not succeed in producing other black holes solutions. If the
cosmological constant is zero, and $W$ is a torus then we can show
that there are no solutions satisfying (i)-(iv). Probably, this result
can also be understood in terms of the classical theorems relating the
existence of horizons and the topology of spacetime. For $k=2$ we have
found black hole solutions with a standard Kruskal picture but new
topologies for $W$ are allowed. These solutions are described below.     

We shall leave for future investigations the problem of solving the
ansatz (\ref{kruskal}) in alternative theories of gravity like Stringy
Gravity or Chern-Simons gravity.

The Einstein equations for a metric of the form (\ref{kruskal}) are
consistent only if the Ricci tensor associated to  $g_{ij}$ satisfy
\begin{equation}
R^{i}_{j} = \zeta\delta^{i}_{j} \ \ ,
\label{einstein}
\end{equation}
where $\zeta$ can be chosen, without lost of generality, to take the
values $0$ or $\pm 1$. This means that,  in general, the Euclidean
space $W$ can be any Euclidean, compact Einstein space with
cosmological constant $(D-k-2)\zeta/2$. In the Schwarzschild case, for
example,  this space is a sphere ($ \zeta=1$). Another example is
provided by the new black holes found in \cite{O},  where $\zeta=-1$
and $W$ is a Riemann surface of  genus $g>1$. 

The Einstein equations for (\ref{kruskal}) are therefore,
\begin{eqnarray}
0 &=&     4(D-k)H^2r'\left[ (k-1)r + (D-k-1)s r' \right]\label{eq2} \\ 
  &+&  2(k-1)HH'r \left[ (k-2)r + 2(D-k)sr' \right]  \nonumber\\
  &+& \left(2\Lambda r^2-(D-k)\zeta\right) H^3 +
(k-1)(k-2)sr^2 H'^{2} \nonumber\\
0 &=& \frac{r(k-2)}{D-k}\left[   3H'^2 - 2HH'' \right] +
4 H(H'r'-Hr'')\nonumber.
\end{eqnarray}

\subsection{Solutions with no cosmological constant}

For  $\Lambda=0$ and $\zeta=0$ ($D-k$ torus), the above equations can
be integrated once giving
\begin{equation}
r^{2(D-k-1)}s^k r'^{2} H^{k-2} = \alpha  ,
\end{equation}
where $\alpha$ is an integration constant. This implies that there is
no regular solution in the horizon $s=0$ for the variables $r$ and
$H$. Therefore, it is not possible to construct a well behaved black
hole geometry with topology $M^k \times T^{(D-k)}$, at least without
the cosmological constant. 
       
For $k=2$ and $\zeta\neq 0$ we have the standard $2$--dimensional
Kruskal diagrams. The general solution, with the regularity conditions
explained
before is given by
\begin{eqnarray}
H &=& \frac{4r_{+} r'}{\zeta} \ \ , \label{H} \\
s
&=&\alpha\exp\left[\frac{(D-3)}{r_{+}}\left(r+\int\frac{dr}{(r/r_{+})^
{D-3}
- 1}\right)\right]  ,
\label{s}
\end{eqnarray} 
where $\alpha$ and $r_+$ are real constants.

The fact that $H$ should be positive and $r$ crescent force $\zeta$
and $r_{+}$ to have the same sign.  From (\ref{s}) we learn that  they
must be positive. The family of solutions obtained can be written in
Schwarzschild coordinates performing the transformation of
coordinates:
\begin{eqnarray}
x^{0} &=& \sqrt{s} \cosh \left(\frac{\sqrt{\zeta(D-3)}}{2r_{+}}t
\right) \
\\ 
x^{1} &=& \sqrt{s} \sinh \left(\frac{\sqrt{\zeta(D-3)}}{2r_{+}}t
\right) \,,
\end{eqnarray}
that brings the metric into the form
\begin{equation}
ds^2 = - N^2 dt^2  +
\frac{(D-3) }{\zeta }\frac{dr^2}{N^2} +r^2 d\Omega^2,
\label{sch}
\end{equation}
with 
\begin{equation}
N^2 = 1-\left[\frac{r_{+}}{r}\right]^{D-3} \ .
\end{equation}
[For $s<0$ we replace $\sqrt{s}$ with $\sqrt{-s}$.  In the
above formulas $s$ is the  function of $r$ given by (\ref{s}).]

If we choose $d\Omega^2$ to be a $(D-2)$--sphere of radius $1$, it is
easy to show that $\zeta=D-3$ and we obtain the Schwarzschild
solution.  

It is possible, nevertheless, to find some new solutions. In
$D=6$,  for example, we can  construct a black hole with topology 
$M^2\times S^2 \times S^2$, which reads
\begin{eqnarray}
ds^2 &=& -N^2 dt^2
+N^{-2} dr^2\\ \nonumber  
&+& \frac{1}{3}r^2\left(d\theta_1^2+\sin^2\theta_{1}d\varphi_1^2+
d\theta_2^2+\sin^2\theta_{2}d\varphi_2^2\right) .
\end{eqnarray}     
Note that for $r_{+}=0$ this metric does not represent a flat
spacetime and it has a naked curvature singularity at $r=0$. 

Similar solutions can be constructed in any dimension, where the
topology of the horizon can be any product of spheres $S^{m_{1}}\times
S^{m_{2}}\times  \cdots \times S^{m_{M}}$, where $m_{i} \ge 2$ and
$\sum_{i} m_{i}=D-2$.

\subsection{Solutions with cosmological constant}

Let us now discuss some solutions  of equations (\ref{eq2})  in presence of cosmological constant.  We will again set
$k=2$, $D>3$ and proceed as in the preceding to obtain a
one--parameter family of solutions given, in Schwarzschild coordinates
by
\begin{equation}
ds^2=-\frac{N^2}{\zeta} dt^2 +
\frac{(D-3)}{\zeta}
N^{-2}  dr^2+ r^2 d\Omega^2  ,
\label{cosmological}
\end{equation}
with
\begin{equation}
N^2 = 1- \frac{\omega}{r^{d-3}}-\frac{\tilde{\Lambda}}{\zeta}r^2,
\end{equation}
$\tilde{\Lambda}=\frac{2\Lambda(D-3)}{(D-1)(D-2)}$, 
and $\omega$ is a real integration constant which is proportional to
the mass. 
Note that we can always eliminate the factor $1/|\zeta|$ in $dt^2$
(but not
the sign of $\zeta$), redefining the time coordinate. Moreover, we can
redefine $r$ and eliminate 
$(D-3)/|\zeta|$ from $dr^2$ (the price we pay here is the redefinition
of
$d\Omega^2$). 

For positive mass and negative cosmological constant we have two
possibilities depending on the sign of $\zeta$. When $\zeta <0$ we
obtain
the generalization of the black hole solutions found in \cite{O,tbh}
for any dimension. Again, when $D>4$, the metric $d\Omega^2$ can
represent any surface with $R^{i}_{j}= \zeta\delta^{i}_{j}$, which has
more solutions than a surface of genus $g$. In $D=6$, for example this
could have the topology $T^2\times T^2$. 

When $\zeta>0$ we obtain the
generalization of Schwarzschild--Anti--de Sitter solutions to higher
dimensions. Again we can extend this solutions to exotic topologies.

For positive cosmological constant and $\zeta>0$ we have the
generalization
of the Schwarzschild--de Sitter  solutions, which have an event
horizon and
a cosmological horizon. When  $\zeta<0$ we have again the same kind of
solution, but with surfaces of constant radius of hyperbolic geometry. 
 
\acknowledgments
During this work we have benefited from many discussions and useful comments raised by Marc Henneaux, Miguel Ortiz, Peter Peld\'an, Claudio Teitelboim and  Jorge Zanelli. This work was partially supported by the grants Nos. 1970150, 3960008 and 3970004 from FONDECYT (Chile). The institutional support to the Centro de Estudios Cient\'{\i}ficos de
Santiago of Fuerza A\'erea de Chile and a group of Chilean companies
(Empresas CMPC, CGE, Codelco, Copec, Minera Collahuasi,  Minera
Escondida, Novagas, Business Design Associates, Xerox Chile) is also
recognized.


\begin{references}

\bibitem{Abbot-Deser} L. Abbot and S. Deser, Nucl. Phys. {\bf B195},
76 (1982).     

\bibitem{Fronsdal} C. Frondsal, Rev. Mod. Phys. {\bf 37}, 221 (1965).

\bibitem{Gibbons} G. Gibbons, in ``Proceedings of Supersymmetry,
supergravity and related topics", del Aguila, de Azcarraga and
Iba\~nez eds, World Scientific 1995.

\bibitem{Maldacena} J.M. Maldacena, hep-th/9711200.

\bibitem{Witten98} E. Witten, hep-th/9802150

\bibitem{Witten89} E.~Witten, Commun.\ Math.\ Phys.\ {\bf 121}, 351
(1989).

\bibitem{Carlip} S. Carlip, {\em Phys. Rev.} {\bf D51}, 632-637
(1995)  

\bibitem{Strominger} A.Strominger, hep-th/9712251

\bibitem{bbo} M. Ba\~nados, T.Brotz and M. Ortiz, hep-th/9802076

\bibitem{BTZ} M. Ba\~nados, C. Teitelboim and J.Zanelli,
{\em Phys. Rev. Lett.} {\bf 69}, 1849 (1992).

\bibitem{BHTZ} M. Ba\~nados, M. Henneaux, C. Teitelboim and
J.Zanelli, Phys. Rev. {\bf D48}, 1506 (1993).

\bibitem{O}  S Aminneborg, I Bengtsson, S Holst and P Peld\'an,  
Class. Quant. Grav. {\bf  13}, 2707 (1996). 

\bibitem{BTZu}  M. Ba\~nados, C. Teitelboim and J.Zanelli, ``New
topology and signature in higher dimensions", unpublished.

\bibitem{b3}  M. Ba\~nados,  Phys. Rev. {\bf D57}, 1068 (1998).

\bibitem{tbh} R. B. Mann, gr-qc/9705007; W. L. Smith and R. B. Mann,
gr-qc/9703007;  J. Creighton and R. B. Mann, gr-qc/9710042; 
D. Brill, J. Louko and P. Peld\'an, Phys. Rev. {\bf D56}, 3600 (1997);
L. Vanzo, Phys.Rev. {\bf D56}, 6475 (1997); R. B. Mann,
gr-qc/9709039. 

\bibitem{Holst-Peldan} S. Holst and P. Peld\'an, Class. Quantum Grav.
{\bf 14}, 3433 (1997).

\bibitem{DJt} S. Deser, R. Jackiw and G. t' Hooft, Ann. Phys. (NY)
{\bf 152}, 220 (1984). 

\bibitem{M-S} J. Maldacena and A. Strominger, hep-th/9804085

\bibitem{Chamseddine} A.H. Chamseddine, 
Nucl. Phys. {\bf B346}, 213 (1990).

\bibitem{Achucarro} A. Ach\'ucarro and P.K. Townsend,  Phys. Lett.
{\bf B180}, 89 (1986).

\bibitem{Moore} G. Moore and N. Seiberg, {\em Phys. Lett.}
{\bf B220}, 422 (1989); S. Elitzur, G. Moore, A. Schwimmer and
N. Seiberg {\em Nucl. Phys.} {\bf B326}, 108 (1989).

\bibitem{Bal} See A. P. Balachandran, G. Bimonti, K.S.
Gupta, A. Stern, {\em Int. Jour. Mod. Phys.} {\bf A7}, 4655 (1992).

\bibitem{B} M. Ba\~nados, {\em Phys. Rev.} {\bf D52}, 5816 (1995).

\bibitem{BGH2} M. Ba\~nados, L.J. Garay and M. Henneaux, Nucl.
Phys. {\bf B476}, 611 (1996); Phys. Rev. {\bf D53}, R593
(1996). 

\bibitem{Nair} V.P. Nair and J. Schiff, Phys. Lett. {\bf
B246}, 423 (1990);  Nucl. Phys. {\bf B371},  329 (1992).

\bibitem{B-Tr-Z}  M. Ba\~nados, R. Troncoso and J.Zanelli, 
Phys.Rev. {\bf D54}, 2605 (1996). 

\bibitem{BTZ2} M. Ba\~nados, C. Teitelboim and J.Zanelli,
{\em Phys. Rev. } {\bf D49}, 975 (1994).

\bibitem{Tr-Z} R. Troncoso and  J. Zanelli, hep-th/9710180. 

\bibitem{CLM}  D. Cangemi, M. Leblanc and R.B. Mann,
{\em Phys. Rev. } {\bf D48}, 3606 (1993).

\bibitem{CH} O. Coussaert and M. Henneaux,  
Phys. Rev. Lett. {\bf 72}, 183 (1994). 



\end{references}
\end{document}